\documentclass[aps,prl,twocolumn,showpacs,superscriptaddress,groupedaddress]{revtex4}  
\usepackage{graphicx}  
\usepackage{dcolumn}   
\usepackage{bm}        
\usepackage{amssymb}   

\begin{document}



\title{Imperfect World of $\beta\beta$-decay Nuclear Data Sets?}
\author{B. Pritychenko}
\email{pritychenko@bnl.gov}
\affiliation{%
National Nuclear Data Center, Brookhaven National Laboratory, Upton, NY 11973-5000, USA 
}

\date{\today}

\begin{abstract}
The precision of double-beta ($\beta\beta$) decay  experimental half-lives and their uncertainties is reevaluated.   
A complementary  analysis of the decay uncertainties  indicates deficiencies due to small size of statistical samples, and incomplete 
collection of experimental information.   Further experimental and theoretical efforts would lead toward more precise values
 of  $\beta\beta$-decay half-lives and nuclear matrix elements.
 \end{abstract}

\pacs{23.40.-s,21.10.-k,28.20.Ka,29.87.+g}
\maketitle

Double-beta  decay was proposed by M. Goeppert-Mayer \cite{35Goe}  as a nuclear disintegration with simultaneous emission of 
 two electrons and two neutrinos
\begin{equation}
\label{myeq.1a}  
(Z,A) \rightarrow (Z+2,A) + 2 e^{-} + 2\bar{\nu}_{e}.
\end{equation}

 There are several double-beta decay processes: $2 \beta^{-}$,  $2 \beta^{+}$, $\epsilon$ $\beta^{+}$, 2$\epsilon$ and 
  possible decay modes:  two-neutrino (2$\nu$), neutrinoless (0$\nu$) and Majoron emission ($ \chi^{0}$)
 \begin{equation}
\label{myeq.1b}  
(Z,A) \rightarrow (Z\pm2,A) + (2 e^{\pm}) + (2\bar{\nu}_{e},\  2{\nu}_{e}\ or\  \chi^{0}).
\end{equation}

The $\beta\beta$-process has been extensively investigated  in the last 80 years \cite{92Boe,02Tre}. These efforts have led to 
 observations of the two-neutrino decay mode and deduction of  decay half-lives. It is the rarest presently-observed 
 nuclear decay. In a recent  analysis  of $\beta\beta$-decay data,  A.S. Barabash \cite{10Bar} claimed that we 
 could deduce precise values of experimental half-lives and extract the corresponding nuclear matrix elements (NME). 
 Unfortunately, these claims are  premature and not very beneficial for the field. The main goal of this work is to 
 investigate the present status of $\beta\beta$-decay research, reanalyze the data, and produce realistic assessments 
 and recommendations that could accelerate  overall progress.

\begin{figure}[htb]
\includegraphics[width=9.5cm, angle=0]{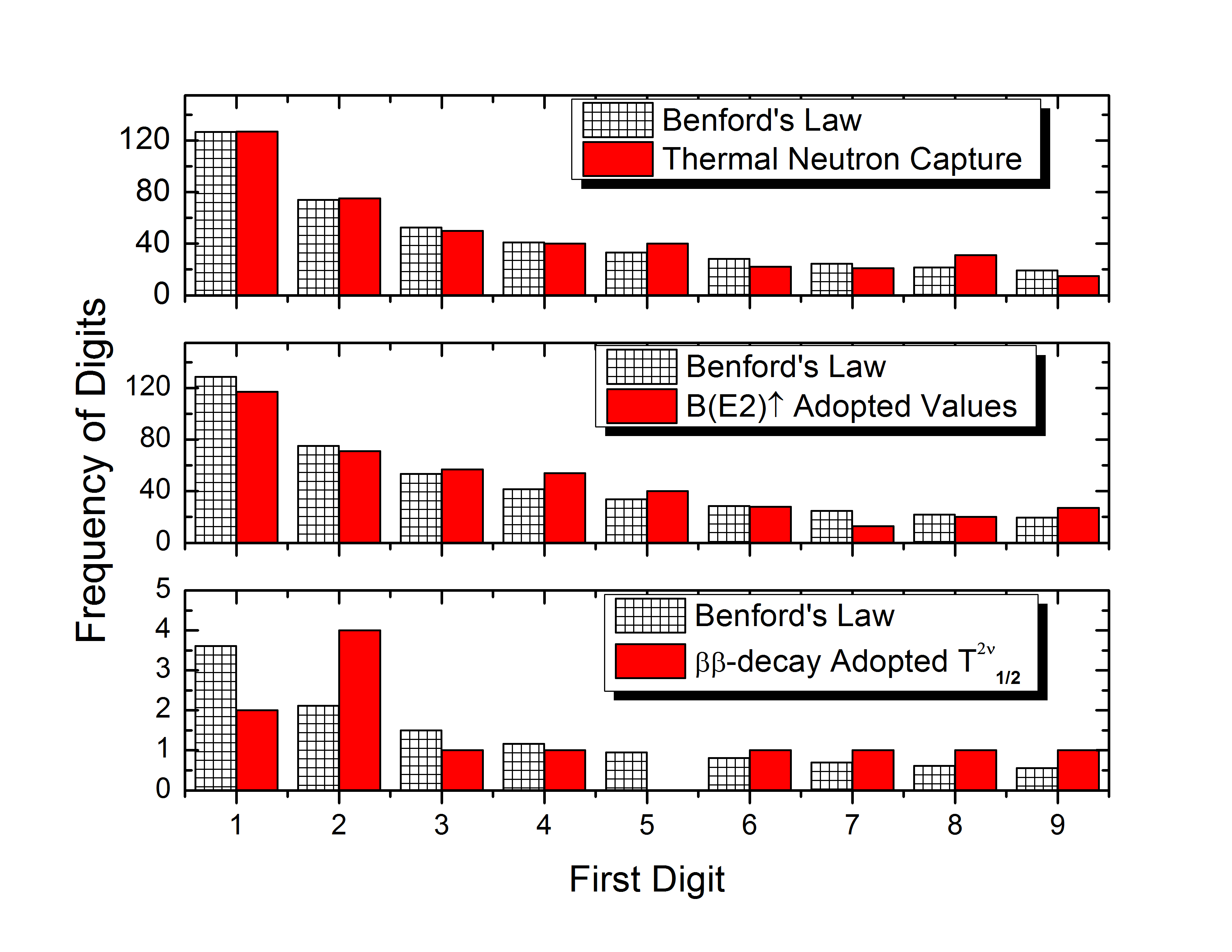}
\caption{Benford's distributions for adopted values of thermal neutron capture cross sections, B(E2)$\uparrow$ and $\beta\beta$-decay T$_{1/2}^{2\nu}$ data sets \cite{06Mug,BE2,14Pri}.}
\label{fig1}
\end{figure}

The $\beta\beta$-decay T$_{1/2}^{2\nu}$ are available from multiple sources \cite{10Bar,14Pri}. A brief summary of  the 
recent National Nuclear Data Center (NNDC) evaluated or adopted values is shown in Table \ref{table1} and plotted in the 
lower part of  Fig. \ref{fig1}.  This plot contains 12 adopted half-lives for nuclei of practical interest.
 The NNDC evaluation is completely based on standard U.S. Nuclear Data Program procedures, and its validity has 
 been extensively scrutinized using theoretical predictions and nuclear systematics arguments \cite{14Pri}. 

\begin{table}[!htb]
\centering
\caption{Adopted $\beta\beta$-decay T$_{1/2}^{2\nu}$ for  0$^{+}$ $\rightarrow$ 0$^{+}$ transitions.  Data are taken from the Ref. \cite{14Pri}.}
\begin{tabular}{c|c|c}
\hline
\hline
Parent nuclide  &    {\it T}$_{1/2}^{2 \nu}$(y) &  {\it T}$_{1/2}^{2 \nu + 0 \nu + \chi^{0}}$(y)  \\
\hline

$^{48}$Ca &  (4.39$\pm$0.58)x10$^{19}$ & \\ 		
$^{76}$Ge &    (1.43$\pm$0.53)x10$^{21}$ &  \\ 		
$^{82}$Se &     (9.19$\pm$0.76)x10$^{19}$ & \\ 		
$^{96}$Zr &    (2.16$\pm$0.26)x10$^{19}$ &  \\ 		
$^{100}$Mo &    (6.98$\pm$0.44)x10$^{18}$ &  \\ 		
$^{116}$Cd &     (2.89$\pm$0.25)x10$^{19}$ &  \\ 		
$^{128}$Te &    & (3.49$\pm$1.99)x10$^{24}$ \\
$^{130}$Te &   (7.14$\pm$1.04)x10$^{20}$ &  \\ 	
$^{136}$Xe &     (2.34$\pm$0.13)x10$^{21}$ &    \\	
$^{130}$Ba &   & (1.40$\pm$0.80)x10$^{21}$ \\
$^{150}$Nd &    (8.37$\pm$0.45)x10$^{18}$ & \\ 		
$^{238}$U  &     & (2.00$\pm$0.60)x10$^{21}$ \\
\hline
\hline
\end{tabular}
\label{table1}
\end{table}

To gain a complementary insight on the statistical sample size, as shown in the Table \ref{table1}, I will resort to a non-traditional approach and consider  
 Benford's Law  \cite{Ben}.  
Figure \ref{fig1} illustrates that relatively large statistical samples for thermal neutron capture cross sections 
and B(E2)$\uparrow$ adopted values of 421 and 427 nuclei \cite{06Mug,BE2}, respectively, are in  
agreement with the law.  The Pearson's cumulative test statistic values  
for nuclear reaction and structure data sets are 8.584 and 15.0, respectively. These cumulative values could be compared with   
upper-tail critical values of $\chi^2$ distribution \cite{NIST1} with 8 degrees of  freedom. 
A similar analysis for the adopted values of $\beta\beta$-decay T$_{1/2}^{2\nu}$ \cite {14Pri} is not possible because one is supposed to have at 
least 5 counts in each  of the nine bins \cite{NIST2}, or the number of $\beta\beta$-decay observations has to quadruple.  
Consequently, there is no sense in applying it to the presently-available $\beta\beta$-decay half-lives due  to the limited number of observed transitions and large experimental uncertainties.

From the dawn of the $\beta\beta$-decay era researchers knew the importance of complete experiments when both released energy 
and angular distribution of decay products have been recorded. The first direct observation of two-neutrino decay mode in 
$^{82}$Se \cite{82Se} and subsequent observations in $^{150}$Nd and $^{48}$Ca \cite{150Nd,48Ca} were made using such techniques. 
These discoveries employed the time projection chambers (TPC) that contained small amounts of target material and, consequently, 
generated limited statistics. The experiments were very complex; however, observation of two-electron events has provided
 clear evidence of the double-beta decay process.

\begin{figure}[!ht]
 \begin{center}
   \begin{tabular}{c}
    \includegraphics[width=8.5cm, angle=0]{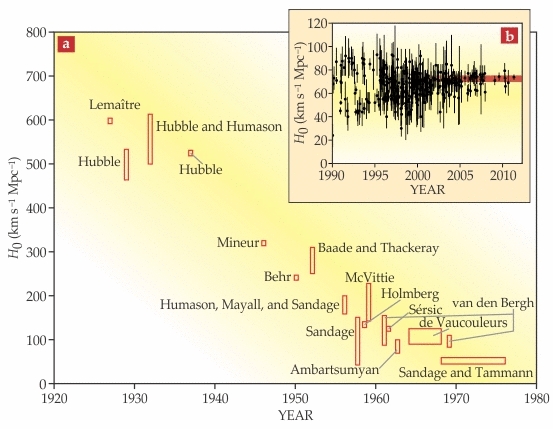} \\
    \includegraphics[width=8cm, angle=0]{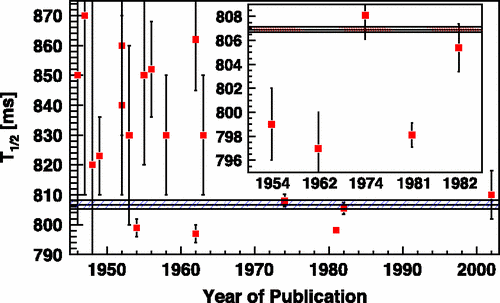} \\
    \includegraphics[width=9cm, angle=0]{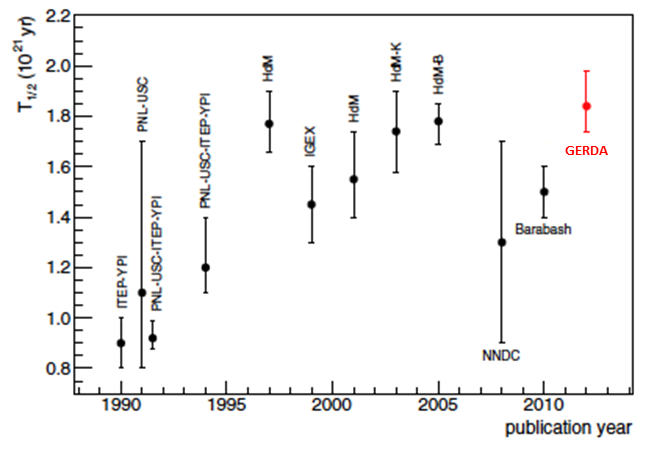} \\
\end{tabular}
  \caption{Chronological record of three different groups of measurements: Hubble constant \cite{13PT}, T$_{1/2}$ of $^{6}$He($\beta$) \cite{12Kne} 
  and $^{76}$Ge$^{2\nu}$($\beta\beta$) \cite{13Ger}. These graphs are courtesy of the cited publications.}
\label{fig2}
\end{center}
\end{figure}

Further developments have progressed using advanced commercially-available detectors  \cite{13Ger,13Maj} and 
large quantities of enriched isotopes. Unfortunately, the usage of commercial detectors often leads to incomplete 
experiments when only energy release information has been collected. To illustrate this point, we should consider 
$\beta\beta$-decay search in $^{76}$Ge. The lower part of Fig. \ref{fig2} depicts a chronological record of the 
measurements \cite{13Ger}. The different half-life values with large uncertainties for two-neutrino mode of decay are 
often explained by the relatively high background in the earlier experiments. 
These experiments were based on an essentially-single source of enriched $^{76}$Ge isotope and slightly different detector 
fabrication and shielding technologies. All of these measurements relied heavily on an excellent energy resolution of Ge 
detectors and suffered from the lack of electron tracking information. Consequently, a final $\beta\beta$-decay spectrum could 
contain the contribution of single-electron events. High energy resolution  is absolutely essential for observations of 
neutrinoless of $\beta\beta$-decay, when decay will produce a sharp peak corresponding to a Q-value between  parent and 
daughter nuclei. Unfortunately, the quantum world is very diverse, and background processes may affect the results. 
It has been demonstrated recently that $\gamma$-ray transitions in Pb and $^{76}$Ge could produce a 2039 keV signal and obscure 
the decay signature \cite{76GePb,76Ge76Ge}.

Electron tracking information is crucial in  $\beta\beta$-decay research. An observation of two-electron events in addition to 
energy release information would help to suppress the radioactive background and decisively prove  $\beta\beta$-decay process 
in a particular nucleus. Perhaps, in addition to HPGe technologies, it could be interesting to 
remeasure $^{76}$Ge T$_{1/2}^{2\nu}$ in a NEMO experiment  that is based on a more extensive set of observables \cite{NEMO}, or explore germanium tetrafluoride (GeF$_{4}$). 
Similar scenarios have been developing with  $\beta\beta$-decay search in $^{136}$Xe \cite{136Xe,136Xe1,136Xe2}. 

In light of this disclosure, it becomes clear that excellent energy resolution of HPGe detectors alone may not be 
sufficient for an observation of neutrinoless mode in $^{76}$Ge. Therefore, the preset day state-of-the-art GERDA and 
Majorana experiments \cite{13Ger,13Maj} may not be able to provide an ultimate answer due to the incomplete nature of 
the data recording.

Analysis  of the lower part of Fig. \ref{fig2} shows that the ``precise"  T$_{1/2}^{2\nu}$ value of 
Barabash \cite{10Bar} strongly contradicts  the latest result of the GERDA experiment   \cite{13Ger}.   
This example shows that it is too early, at this point,  to state  a high precision of adopted half-lives and NME, and 
large error bars of the previous NNDC evaluation \cite{10Pri}  are more appropriate. All NNDC  values were produced from 
the experimental half-lives when new measurements would trigger the data reevaluation every 5-7 years. 

A complementary analysis of the upper and lower parts of Fig. \ref{fig2} shows that the presently-discussed 
situation with $\beta\beta$-decay measurements is not unique in physics. It is rather a common occurrence when initial, 
pioneering measurements are not very accurate and often discrepant. It definitely has happened 
to  Hubble constant and $^{6}$He $\beta$-decay experiments \cite{13PT,12Kne}, as  shown in the upper and middle part of 
Fig. \ref{fig2}. This figure content is  based on  original graphs borrowed from the Refs. \cite{13Ger,13PT,12Kne}. 
Other recent cases of inconsistent data include the discrepant  decay scheme and half-life of $^{139}$Ba \cite{12Dan}, 
and cross section  values. In the recent review of neutron cross section deficiencies, M.B. Chadwick \cite{15Chad} compiles 
an impressive list that includes  $^{235}$U and $^{197}$Au fast neutron capture cross sections. 
The last cross section value was used in calibration of the stellar nucleosynthesis KaDONiS database \cite{06Kad} and has a 
broad impact across  neutron physics. We are surrounded by a large number of imperfect nuclear data sets and constantly 
work on their improvement.

The experimental  $\beta$$\beta$(2$\nu$)-decay half-lives include contributions from the nuclear structure effects and  
decay kinematics.  $T_{1/2}^{2\nu}$ values are often described as follows 
 \begin{equation}
\label{myeq.Half}
 \frac{1}{T_{1/2}^{2\nu} (0^{+} \rightarrow 0^{+})} = G^{2 \nu} (E,Z)  |M^{2 \nu}_{GT} - \frac{g^{2}_{V}}{g^{2}_{A}} M^{2 \nu}_{F}|^{2}, 
 \end{equation}
 where the function  $G^{2 \nu} (E,Z)$ results from lepton phase space integration and contains all the relevant 
 constants \cite{92Boe}.  Equation (\ref{myeq.Half}) highlights a direct  dependency of experimental NME on the calculated 
 values of phase space factors.

 Both Barabash's and NNDC's evaluated NME were based on the best available phase space factor calculations in 2010 and 2013, 
 respectively \cite{92Boe,98Suh,12Kot}. Table \ref{table3} shows   the evolving   values of $^{76}$Ge phase space factors, 
 and the table data raise a question about calculation limitations and possible model dependency of the PSFs. 
 The observed discrepancies between recent calculations of Kotila \& Iachello \cite{12Kot} and  Stoica \& Mirea \cite{13St} 
 create  reason for concern. The exact values of phase space factors are needed in order to deduce the precise values 
 of experimental NME for comparison with recent theoretical calculations of Senkov and Horoi \cite{14Sen}. It will be highly 
 beneficial for the field if a third group of theorists will clarify the situation. 
 The formula (\ref{myeq.Half}) gets even more complex for the neutrinoless mode where half-life also depends on a neutrino mass, 
 and neither neutrino mass or NME can be easily disentangled.
 \begin{table}
\centering
\caption{ $^{76}$Ge $\beta\beta$-decay phase space factor (PSF) values in 10$^{-21}$ yr$^{-1}$ for $0^{+} \rightarrow 0^{+}$  transition.}
\begin{tabular}{c|c|c}
\hline
\hline
Authors & Year & PSF \\
\hline
Boem \& Vogel \cite{92Boe}  & 1987 & 130.54 \\
Doi  {\it et al.} \cite{93 Doi} &  1993 & 53.8 \\
Suhonen \& Civitarese \cite{98Suh} & 1998 & 52.6 \\
 Kotila \& Iachello \cite{12Kot} & 2012 & 48.17 \\
Stoica \& Mirea \cite{13St} &  2013 & 43.9 \\
\hline
\hline
\end{tabular}
\label{table3}
\end{table} 

Finally, the double-beta decay experimental half-lives have been reanalyzed using  standard nuclear data techniques.  
The analysis of the two-neutrino mode of $\beta\beta$-decay   data sets indicates  their uncertainties are high due to relatively 
small experimental statistics and incomplete collection of decay information.    We live in an era of $\beta\beta$-decay observations with large uncertainties. 
It may take years before future complete or ``almost" complete experiments like low-pressure TPC \cite{82Se,150Nd,48Ca} or  
simultaneous measurement of positrons and $\gamma$-rays \cite{78Kr}  will improve the data. The precision of 
experimental NME strongly depends on the quality of phase space factors, and  additional  theoretical calculations are 
necessary to clarify  the discrepant results. 

Double-beta has been an exciting nuclear physics phenomenon that has slowly revealed its experimental signatures in the 
last 30 years. It may take  many painstaking experimental and theoretical efforts before the process will be well 
understood and measured. It is not an unusual situation in a history of science; in fact, it is a very common case. 
Therefore, one can assume that it may take another 30 years before all decay modes, experimental half-lives and NME 
values will be finalized.

The author is indebted to Dr. M. Herman (BNL) for support of this project  and grateful to Dr. V. Unferth (Viterbo University) 
for help with the manuscript. This work was funded by the Office of Nuclear Physics, Office of Science of the U.S. Department 
of Energy, under Contract No. DE-AC02-98CH10886 with Brookhaven Science Associates, LC.

\end{document}